\documentclass[a4paper]{jpconf}
\usepackage[utf8]{inputenc}
\usepackage{graphicx}
\usepackage{subfigure}
\usepackage{multirow}
\usepackage{amsmath}
\usepackage{mathtools}
\usepackage{times}
\usepackage{xcolor}
\usepackage{float}

\begin{document}
\title{Mobility of localized beams in non-homogeneous photonic lattices}
\author{M A Sabogal, I C Parra, M Bandera, J Gallardo and Cristian Mejía-Cortés}
\address{Programa de Física, Facultad de Ciencias Básicas, Universidad del Atlántico, Puerto Colombia 081007, Colombia.}
\ead{ccmejia@googlemail.com}

\begin{abstract}
Waveguide arrays offer enormous potential to design circuit elements essential
to fabricate optical devices capable to processing information codified by
light. In this work we study the existence and stability of localized beams 
in one dimensional photonic lattices composed by a Kerr type waveguide 
array. We analyzed the case where discrete translation symmetry is 
broken in as much as one of the waveguides lacks a nonlinear response. 
Specifically, we determined the space of parameters where a coherent 
and robust mobility across the lattice is achieved. Moreover, we 
calculate the reflection and transmission coefficients when localized 
beams interact directly with the impurity, finding that it behaves as
a variable filter depending of system parameters.  Our results would
shed light on develop solutions to keep unaltered information
during its transmission within future optical devices. 

\end{abstract}

\section{Introduction}


Localized modes have been studied extensively since middle of last
century~\cite{PhysRev.109.1492}. Many physical systems exhibit different
phenomena which lead to formation of this kind of excitation. For example, the
first system where light localization was predicted and observed was in optical
fibers~\cite{doi:10.1063/1.1654836}. Here, light pulses were able to travel long
distances without distortion, due to a balance between nonlinear response of
material and chromatic dispersion of light~\cite{MO35250} .  This kind of pulse
received the name of {\it optical soliton} and since then they have been widely
used in telecommunications~\cite{Segovia_Cabrera_2015}.

Optical periodic systems have attracted enormous attention during last three decades
because they bear enormous potential in technological applications. Their
underlying characteristics offer the possibility of manage the light behaviour
either over long distances or short ones. For example, at big scale photonic
crystal fibers, optical fibers with a micro structured cross
sections~\cite{Russell:06, Russell358}, can be employed in fiber-optics
communications~\cite{Roberts:05}, but also they can be used as sensors with high
resolution~\cite{Cregan1537}. They also offer the possibility to manage light
propagation in short scale. Logical operations similar to those involved with
electron currents can be mimic in a completely optical ``microprocessor'' or
photonic chip. It can be plausible due to the refractive index in theses
systems possesses a periodical distribution, hence, there are forbidden regions
for light propagation~\cite{Joannopoulos}. Experimentally, photonic lattices has
been implemented by creating waveguide arrays in several media. For example,
by using a femto-second pulsed laser on an amorphous (non-crystalline) phase
silicon glass, it can be possible to ``write'' waveguides by modifying the
nominal refractive index around the area where it is has been
focused~\cite{Szameit_2010}.  Photo-refractive crystals are systems where
these waveguide arrays also can be written by and induction process, due to
its refractive index changes by the light intensity variation, i.e., by a
non-linear response of the electric field~\cite{Armijo:14}. On the other
hand, research on coherent transfer of light stays as a hot topic over the
years due to its direct implications in design technological devices for
controlled transport of information. 

A change in the periodic distribution of the refractive index, by introducing an
impurity into the lattice, results in the scattering of transverse traveling
waves. For example, when the impurity comes from localized solutions, the
scattering of plane waves by them has opened the possibility to observe Fano
resonances~\cite{PhysRev.124.1866}. In a nonlinear optical context, it has been
observed that scattering of solitons in waveguide arrays moving towards impurity
potentials, has a complex phenomenology~\cite{PhysRevLett.99.133901}. Moreover,
by adjusting the strength of linear impurities a completely trapping regime can
be tailored~\cite{Morales-Molina:06}. In the present study we address the case
of nonlinear photonic lattices with an embedded linear impurity, which
eventually improves the manipulation of light beam across the lattice. We hope
that our results may be interesting in the design of optical limiters, barriers
and gates for future photonic chips. 

The paper is organized as follows: in Section~\ref{model}, we introduce the
model and develop the main formalism employed to identify nonlinear stationary
solutions, as well as, a favorable domain in terms of system parameters to
achieve coherent and robust mobility. In Section~\ref{modes}, we report findings
on the existence and stability of localized stationary solutions around the
lattice impurity. Section~\ref{mobility} is devoted to estimate the optimal
domain for coherent and robust mobility of nonlinear modes. The analysis on
scattering problem between nonlinear modes and lattice impurity is presented in
Section~\ref{scattering}. Finally, in Section~\ref{conclusions}, we summarize
and draw our main conclusions.

\section{Model}
\label{model}

The Discrete Nonlinear Schrödinger Equation (DNLSE) represents one of the most
important models in nonlinear physics. For example, in classical mechanics, this
equation describes a particular model for a system of coupled anharmonic
oscillators~\cite{eilbeck2003discrete}. On the other hand, this model predict
the existence of localized modes of excitation of Bose–Einstein condensates in
periodic potentials such as those generated by counter-propagating laser beams
in an optical lattice~\cite{Franzosi_2011}. In nonlinear optics, this equation
combines phenomena related to the dispersion and/or diffraction of
electromagnetic waves with those generated by higher order electric polarization
in periodical media~\cite{khare2006discrete}. 

When impurities are introduce to the system the translational symmetry
becomes broken, which leads to the formation of localized modes around the
defects~\cite{kevrekidis2009discrete}. For the case when the effect of
impurity is the lack of nonlinear response in a specific waveguide, we can
model the propagation, along the $\hat z$-axis, of the corresponding electric field amplitude 
present in the $n$-th guide, $E_ {n} (z)$, as a variant of a more general
DNLSE
\begin{equation}
    i\frac{dE_{n}}{dz} + \zeta_{n+1}E_{n+1}+\zeta_{n-1}E_{n-1} +
    \gamma(1-\delta_{n,n_{i}})|E_{n}|^{2}E_{n} = 0,
    \label{eq1}
\end{equation}
where $ i = \sqrt {-1} $ and $\delta_{n,n_{i}}$ is the Kronecker symbol.  The
nonlinear response of the media is represented by the parameter $\gamma$, which
is proportional to the nonlinear refraction index of the medium.  Here we assume
that $\zeta_ {n}$, the coupling between waveguides, is the same for each $n$, i.
e., $\zeta_ {n}= \zeta $. The Equation~(\ref{eq1}) has two conserved quantities,
the generating function that corresponds to the Hamiltonian ($H$)
\begin{equation}
    H = -\sum_{n=1}^{N}\left(\zeta E_{n}(z)E^{\ast}_{n+1}(z)+\text{c.c.} +
    \frac{\gamma}{2}(1-\delta_{n,n_{i}})|E_{n}(z)|^{4}\right),
    \label{eq2}
\end{equation}
where the symbol $^\ast$ and c.c. denote the complex conjugate, and the norm or optical power ($P$) in the system
\begin{equation}
P = \sum_{n=1}^{N}|E_{n}(z)|^{2}.
    \label{eq3}
\end{equation}
It is worth to mention here that these two conserved quantities will be
monitored during all the calculations along this work, because they will serve
to check the validity of our numerical findings.

\section{Families of nonlinear modes}
\label{modes}

We look for stationary solutions of Equation~(\ref {eq1}) in the form $ E_ {n} =
\phi_ {n} \exp {(i \lambda z)} $, where the amplitudes $ \phi_ {n} $ are 
real quantities that satisfy the following system of nonlinear algebraic
equations
\begin{equation}
    -\lambda\phi_{n} + \zeta(\phi_{n+1}+\phi_{n-1}) + \gamma(1-\delta_{n,n_{i}})\phi_n^3=0,
    \label{eq4}
\end{equation}
being $\lambda $ the propagation constant of the stationary solutions.
According on the sign of $\gamma$ the nonlinear effect of the system
can be of the self-focusing type ($\gamma>0$) or self-defocusing type
($\gamma<0$). Throughout the rest of the paper, we assume to deal with 
self-focusing type media.

We solve the model~(\ref{eq4}) by implementing a Newton-Raphson scheme.  We
start from a localized seed around the impurity and in a few number of
iterations the algorithm converges to localized stationary solutions. We check
the stability of localized solutions by performing the standard linear stability
analysis.  From now on we use solid (dashed) lines to denote families with
stable (unstable) solutions.  It is interesting to study the consequences that
the value of the nonlinear constant of the impurity of the lattice entails,
specifically, the type of solutions that exist around the defect and its
mobility around it. Recently,  it has been found that modes centered at site are
the only stable family of solutions that exist around the
impurity~\cite{mejia2019nonlinear}, however, we found that the stability region
of these modes depends on the value of the nonlinearity of impurity.

\begin{figure}[H]
\centering
    \includegraphics[width=1\textwidth]{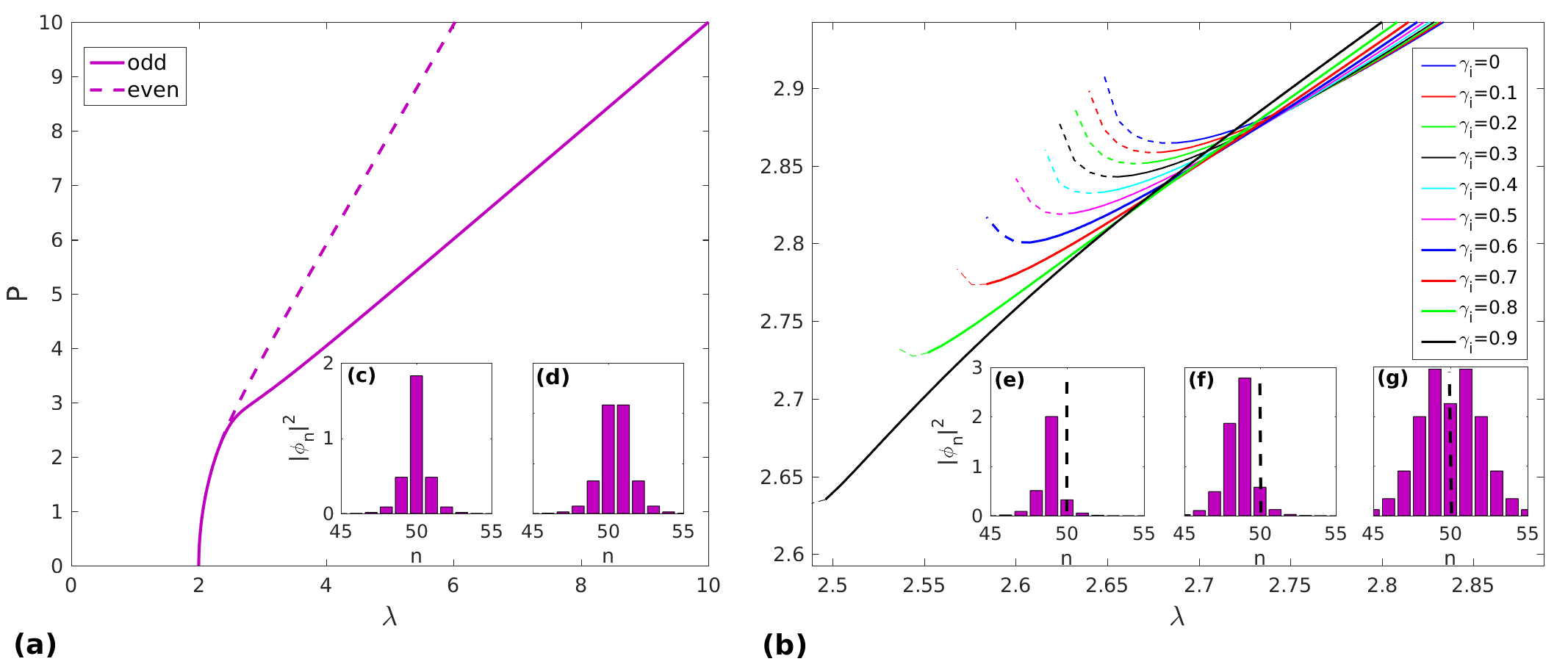}
    \caption{(a) $P$ vs $\lambda$ diagram for odd and even families
    of solutions far away from the impurity. Bottom inset displays 
    both solutions for $ P $ = 3.00  and $\lambda$ = $2.85$ (c) and $2.62$ (d),
    respectively. (b) $P$ vs $\lambda$ diagram for odd families
    around impurity for different values of $\gamma$. Bottom inset displays 
    three different type solutions, near to impurity for $ P $ = 3.00,
    $\gamma=0$ and $\lambda$ = $2.91$ for the odd mode (e), $\lambda$ = $2.66$ 
    for even mode (f) and $\lambda$ = $2.28$ for symmetrical mode (g).}
    \label{fig1}
\end{figure}

Figure~\ref{fig1}(a) shows the families of solution in the space of ($P$,$\lambda$) for
the even and odd modes far from the defect. Families for odd modes around the defect
for ten values of $ \gamma_ {i} $ between $ 0 $ and $ 0.90 $ is displayed at 
Figure~\ref{fig1}(b). It can be seem here that when nonlinearity for defect
diminish there is a reduction in the region of existence and stability of this solutions.
The inset in Figure~\ref{fig1}(a) sketches the odd (c) and even (d) modes 
belonging to the families represented by solid and dashed curves, respectively.
On the other hand, bottom inset at Figure~\ref{fig1}(b) displays three modes
around the impurity that belongs to the odd (e), even (f) and symmetrical (g) 
families of solution, when $\gamma=0$, for three different values of 
$\lambda$. Last two families are not illustrated in this work but they have
been reported recently in reference~\cite{mejia2019nonlinear}.

\section{Mobility of localized modes}
\label{mobility}
\begin{figure}[h!]
\centering
\includegraphics[width=1\textwidth]{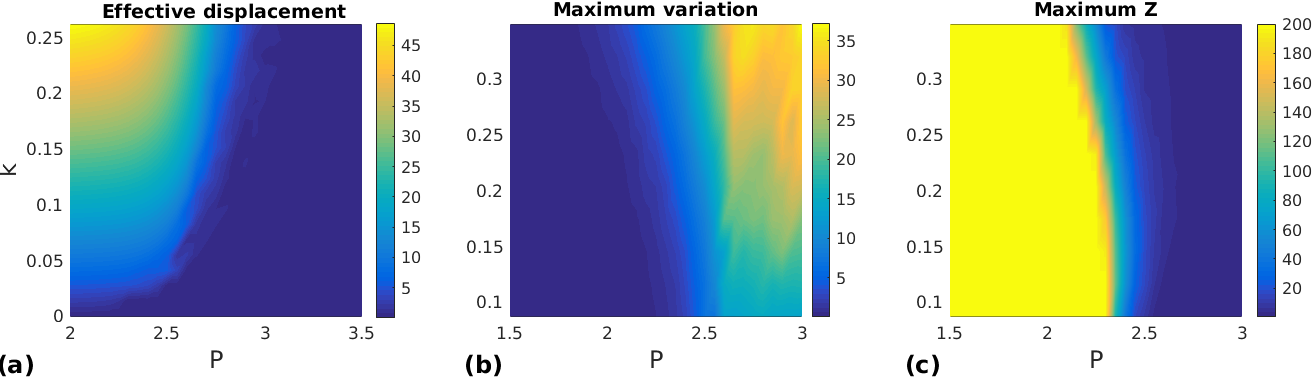}
\caption{Effective displacement of the center of mass (a), maximum variation of
the velocity angle of the center of mass (b) and maximum distance at which change 
in the velocity angle is below $ 10.00 \% $ (c), as function of $P$ and $k$. }
\label{fig2}
\end{figure}

With the aim to study the interaction between solitons and the linear impurity,
it is mandatory to determine the zones in the space $(P,k)$ of powers and
momentum, in which coherent mobility is guaranteed. It is well known that the
mobility of discrete solitons is restricted by the Peierls-Nabarro (PN)
barrier~\cite{kevrekidis2009discrete}, which exists due to the non-integrability
of the DNLSE~\cite{brazhnyi2013interaction, peschel2002optical}. This barrier can
be estimated as the difference in energy ($H$) between the solutions of the
fundamental modes, sketched at top left inset in Figure~\ref{fig1}(a).  By
applying a power constraint to the Newton-Raphson method we could compare the
value of the Hamiltonian (energy) of the modes that have the same
power~\cite{lederer2008discrete}. In that way, we can identify those regions
where $H$ is similar, both for odd and even modes, which implies that these
solutions, previously endowed with momentum, can move across the lattice in an
adiabatic way, i. e., they can transform dynamically into the another one
almost without radiating energy and preserving their shape. 
It is well known that for
the Kerr nonlinearity, there is a critical power where PN barrier is large
enough for the soliton to be confined in the initial guide
\cite{ahufinger2004creation}. In order to determine these zones, solitons with
different configurations of $ k $ and $ P $ were propagated. For the case of
good mobility, their effective displacement was quantified, in terms of their
center of mass $CM \coloneqq \sum_{n = 1}^{N} n |\phi_ {n}|^2/P$, as shown in
Figure~\ref{fig2}(a).

It is clear that for power greater than $P\approx 3.00 $ the mobility of the soliton
is almost zero no matter which was the impinged momentum on the mode. In order
to select optimal parameter domain where coherent mobility is guaranteed we
calculate the maximum variation of the velocity angle of the center of mass,
along the propagation distance with respect to the initial angle.
In Figure~\ref{fig2}(b) it can be observed that for values greater than $ P \approx2.50$
the variation of the initial angle is greater than $ 5.00 $ degrees and independent
of the momentum. We also refine our procedure by identifying the maximum
distance at which the initial angle is conserved, undergoing a change of less
than $ 10.00\% $ for different configurations of $ k $ and $ P $. As shown in
Figure~\ref{fig2}(c), for powers below $ P \approx 2.30 $ the criterion is fulfilled for
the entire propagation distance.  The above suggests that the range $ \mid k
\mid \leq \pi/9 $ and $ P <2.30 $ ensures a coherence mobility of the
information.

\section{Pulse transmission}
\label{scattering}
\begin{figure}[H]
\centering
\includegraphics[width=1\textwidth]{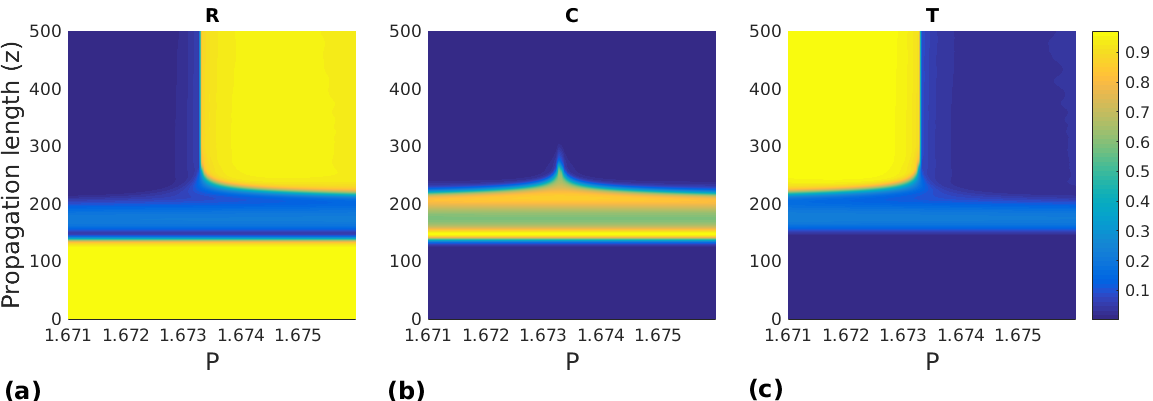}
\caption{Reflection (a), Capture (b) and Transmission (c) coefficients
as function of $P$ and $z$ near of the critical power 
$ P_ {c} = 1.6732 $, for $ k = \pi/18 $, $ \zeta = \gamma = 1.00 $ and $\gamma_{i}=0$.}
\label{fig3}
\end{figure}

Let us now consider the interaction between a discrete soliton, endowed with 
a transverse momentum $k$, and one impurity located at $ n_i = N / 2 $. 
We numerically integrate the model~(\ref{eq1}) with a Runge Kutta scheme by 
taking as initial condition $\phi_n^k=\phi_n\exp{(ikn)}$, being $\phi_n$
a stationary mode of the system. When the soliton reaches the impurity,
the radiation can be reflected, transmitted and/or captured. To analyze
the soliton-impurity interaction we define the reflectance $ (R) $,
transmittance $ (T) $ and the capture fraction $ (C) $ coefficients as
\begin{equation}
    R\coloneqq \frac{\sum_{n=1}^{n_{i}-\Delta}|\phi_{n}|^2}{\sum_{n=1}^{N}|\phi_{n}|^2},  
    \hspace{0.5cm}
    T\coloneqq \frac{\sum_{n=n_{i}+\Delta}^{N}|\phi_{n}|^2}{\sum_{n=1}^{N}|\phi_{n}|^2},    
    \hspace{0.5cm}
    C\coloneqq \frac{\sum_{n_{i}-\Delta}^{n_{i}+\Delta}|\phi_{n}|^2}{\sum_{n=1}^{N}|\phi_{n}|^2}.    
    \label{rtc}    
\end{equation}
Here $ \Delta $ is defined as the size of the impurity. As we are dealing with
conservative systems, it is clear that the condition $ R + T + C = 1 $ must be
fulfilled during beam evolution.  It has been observed that when solitons of the
same power with different $ k $ are sent toward the impurity, there is a
critical momentum $ k_ {c} $ for which the soliton is trapped around the
defect~\cite{mejia2019nonlinear}. Therefore, for values $ k <k_ {c} $ the
solitons are reflected and for $ k> k_ {c} $ they are transmitted. We observe a
similar behavior here, by varying the optical power of the soliton with a fixed
$ k $, which interacts with the impurity. As can be seen from Figure~\ref{fig3}(a),
it is clear that impurity behaves like a filter once the power of modes is near to
$P_c\approx 1.674$, namely, the critical power. Around this value of $P_c$ there exist
a narrow region where light can be trapped by the impurity [cf. Figure~\ref{fig3}(b)] over 
significant distances of propagation. Below this critical power ($ P_ {c} $),
the radiation becomes transmitted almost entirely as is illustrated in Figure~\ref{fig3}(c).

\begin{figure}[H]
\centering
\includegraphics[width=1\textwidth]{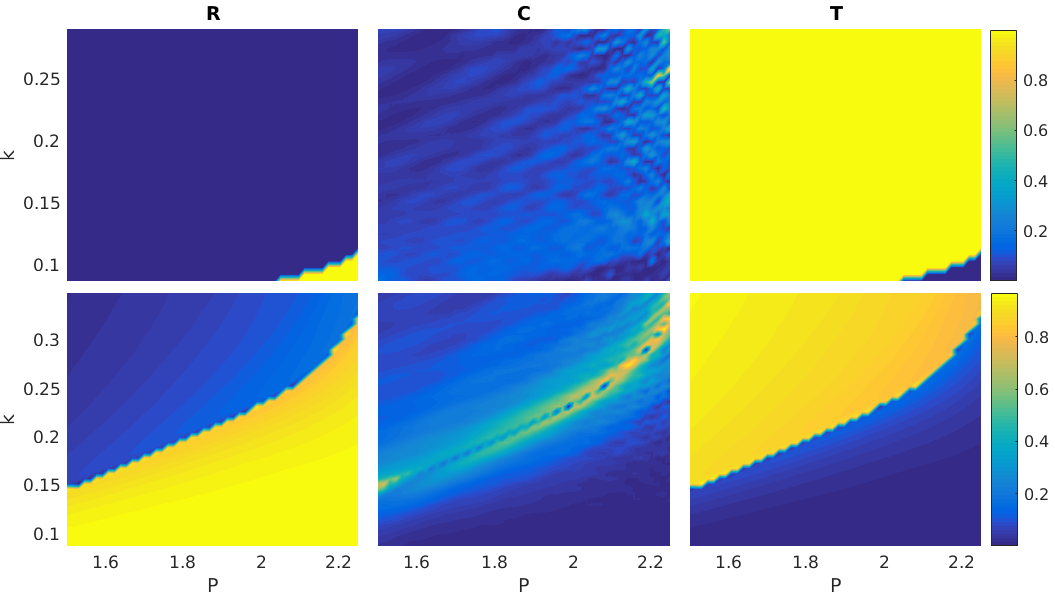}
\caption{Color map of the coefficients (R) Reflectance (C) Capture and (T) 
Transmittance as function of the transverse momentum $ k $ and the optical
power, for $ \zeta = \gamma = 1.00 $ and $\gamma_{i}=0.90$ for the upper row and
$\gamma_{i}=0.00$ for the lower row.}
\label{fig4}
\end{figure}

With the aim to have a more detailed landscape on how the optical power and
the transverse momentum affect the soliton-impurity interaction, we analyze the
propagation of solitons with different configurations of $ k $ and $ P $, moving
towards the defect, for two different values of $\gamma_{i}$. In
order to ensure a consistent estimation of $R$, $C$ and $T$ coefficients, we are going to calculate
coefficients at the longitudinal distance $ z $, after the collision with the
impurity, equal to the longitudinal distance that they had to travel before to
collide with. Figure~(\ref{fig4}) display these coefficients for two values of 
nonlinear parameter; upper row for $\gamma_i=0.90$ and lower row for $\gamma_i=0.00$.
Domains where total reflection $R$ and
transmission $T$ are guaranteed in each case are displayed in first and third column,
respectively. Likewise, sub-spaces where solitons can be trapped around the
impurity correspond with bright spots in middle column. Finally, comparing the two
cases, it is observed that by increasing the value of the nonlinearity of
the defect, a shift or increase in the critical values for which the
impurity behaves like a filter is obtained, approaching the homogeneous
regime when the order between the nonlinear constant of the impurity
and the network tends to one.

\section{Conclusions}
\label{conclusions}
In this work we address the problem of interaction between discrete solitons
and  a linear impurity in a photonic lattice composed by a one dimensional
waveguides array. To do that, we begun calculating the stationary modes that
exist around the defect. The analysis of the mobility of solitons far away
from impurity allow us to determine the areas where the high mobility and 
coherence of information is guaranteed, as function of the transverse momentum
and the optical power. The interaction between the impurity and the nonlinear 
modes displays a corpuscular dynamic. The regions of reflectance and total 
transmittance in which the impurity behaves as an optical limiter were 
determined. Besides, domains for critical power and momentum values in 
which the soliton is trapped around the impurity where identified. Depending 
of the subspace in the parameter space we observe drastic change in the 
reflectance and transmittance coefficients as function of $\gamma_{i}$. 
We hope that these results may be interesting in the design of optical 
limiters for solitons, that allow optimization of the manipulation and 
transmission of information within novel photonic chips. We pretend 
to extend this kind of analysis in systems with more dimensions, as well as,
those ones with exotic dispersion relations.

\section*{References}
\providecommand{\newblock}{}


\begin{thebibliography}{10}
\expandafter\ifx\csname url\endcsname\relax
  \def\url#1{{\tt #1}}\fi
\expandafter\ifx\csname urlprefix\endcsname\relax\def\urlprefix{URL }\fi
\providecommand{\eprint}[2][]{\url{#2}}


\bibitem{PhysRev.109.1492}
Anderson P~W 1958 {\em Phys. Rev.\/} {\bf 109}(5) 1492--1505

\bibitem{doi:10.1063/1.1654836}
Hasegawa A and Tappert F 1973 {\em Applied Physics Letters\/} {\bf 23} 142--144

\bibitem{MO35250}
Ar\'evalo E, Ram\'irez C and Guzm\'an A 1995 {\em MOMENTO\/} {\bf 0}

\bibitem{Segovia_Cabrera_2015}
Segovia F~A and Cabrera E 2015 {\em Redes de Ingenier\'ia\/} {\bf 6} 26--32

\bibitem{Russell:06}
Russell P~S 2006 {\em J. Lightwave Technol.\/} {\bf 24} 4729--4749

\bibitem{Russell358}
Russell P 2003 {\em Science\/} {\bf 299} 358--362

\bibitem{Roberts:05}
Roberts P~J, Couny F, Sabert H, Mangan B~J, Williams D~P, Farr L, Mason M~W,
  Tomlinson A, Birks T~A, Knight J~C and Russell P~S 2005 {\em Opt. Express\/}
  {\bf 13} 236--244

\bibitem{Cregan1537}
Cregan R~F, Mangan B~J, Knight J~C, Birks T~A, Russell P~S~J, Roberts P~J and
  Allan D~C 1999 {\em Science\/} {\bf 285} 1537--1539

\bibitem{Joannopoulos}
Joannopoulos J~D, Villeneuve P~R and Fan S 1997 {\em Nature\/} {\bf 386}
  143--149

\bibitem{Szameit_2010}
Szameit A and Nolte S 2010 {\em Journal of Physics B: Atomic, Molecular and
  Optical Physics\/} {\bf 43} 163001

\bibitem{Armijo:14}
Armijo J, Allio R and Mej\'{i}a-Cort\'{e}s C 2014 {\em Opt. Express\/} {\bf 22}
  20574--20587

\bibitem{PhysRev.124.1866}
Fano U 1961 {\em Phys. Rev.\/} {\bf 124}(6) 1866--1878

\bibitem{PhysRevLett.99.133901}
Linzon Y, Morandotti R, Volatier M, Aimez V, Ares R and Bar-Ad S 2007 {\em
  Phys. Rev. Lett.\/} {\bf 99}(13) 133901

\bibitem{Morales-Molina:06}
Morales-Molina L and Vicencio R~A 2006 {\em Opt. Lett.\/} {\bf 31} 966--968

\bibitem{eilbeck2003discrete}
Eilbeck J~C and Johansson M 2003 The discrete nonlinear schr{\"o}dinger {\em
  Proc. 3rd Conf.: Localization and Energy Transfer in Nonlinear Systems\/}
  p~44

\bibitem{Franzosi_2011}
Franzosi R, Livi R, Oppo G~L and Politi A 2011 {\em Nonlinearity\/} {\bf 24}
  R89--R122

\bibitem{khare2006discrete}
Khare A, Rasmussen K~{\O}, Salerno M, Samuelsen M~R and Saxena A 2006 {\em
  Physical Review E\/} {\bf 74} 016607

\bibitem{kevrekidis2009discrete}
Kevrekidis P~G 2009 {\em The discrete nonlinear Schr{\"o}dinger equation:
  mathematical analysis, numerical computations and physical perspectives\/}
  vol 232 (Springer Science \& Business Media)

\bibitem{mejia2019nonlinear}
Mej{\'\i}a-Cort{\'e}s C, Cardona J, Sukhorukov A~A and Molina M~I 2019 {\em
  Physical Review E\/} {\bf 100} 042214

\bibitem{brazhnyi2013interaction}
Brazhnyi V~A, Jisha C~P and Rodrigues A 2013 {\em Physical Review A\/} {\bf 87}
  013609

\bibitem{peschel2002optical}
Peschel U, Morandotti R, Arnold J~M, Aitchison J~S, Eisenberg H~S, Silberberg
  Y, Pertsch T and Lederer F 2002 {\em JOSA B\/} {\bf 19} 2637--2644

\bibitem{lederer2008discrete}
Lederer F, Stegeman G~I, Christodoulides D~N, Assanto G, Segev M and Silberberg
  Y 2008 {\em Physics Reports\/} {\bf 463} 1--126

\bibitem{ahufinger2004creation}
Ahufinger V, Sanpera A, Pedri P, Santos L and Lewenstein M 2004 {\em Physical
  Review A\/} {\bf 69} 053604

\end{thebibliography}
\end{document}